\documentclass[preprint2]{aastex}

\shorttitle{A blue supergiant star in Virgo cluster}
\shortauthors{Ohyama \& Hota}

\begin{document}


\title{Discovery of a possibly single blue supergiant star in the intra-cluster region of Virgo cluster of galaxies}


\author{Youichi Ohyama\altaffilmark{1} \& Ananda Hota\altaffilmark{2,1}}
\affil{$^1$Institute of Astronomy and Astrophysics, Academia Sinica,
PO Box 23-141, Taipei 106, Taiwan}
\affil{$^2$UM-DAE Centre for Excellence in Basic Sciences, Vidyanagari, Mumbai-400098, India} 



\begin{abstract}
IC~3418 is a dwarf irregular galaxy falling into the Virgo cluster, and a 17~kpc long trail is seen behind the galaxy, 
which is considered to have formed due to ram pressure stripping. The trail contains compact knots and diffuse blobs of 
ultraviolet and blue optical emission and, thus, it is a clear site of recent star formation but in an unusual environment, 
surrounded by a million degree intra-cluster medium. We report on our optical spectroscopy of a compact source in the trail, 
SDSS J122952.66+112227.8, and show that the optical spectrum is dominated by emission from a massive blue supergiant star. 
If confirmed, our report would mark the farthest star with spectroscopic observation. We interpret that a massive O-type 
star formed in situ in the trail has evolved recently out of the main sequence into this blue supergiant phase, and now 
lacks any detectable spectral sign of its associated H~{\sc ii} region. We argue that turbulence within the ram pressure 
striped gaseous trail may play a dominant role for the star formation within such trails.
\end{abstract}


\keywords{galaxies: clusters: individual (Virgo) --- galaxies: individual (IC~3418) --- intergalactic medium --- stars: formation --- supergiants}



\section{Introduction}

As galaxies fall into the huge gravitational potential of clusters of galaxies with high velocity, 
interaction with the hot intra-cluster medium strips off cool gas from the main body of 
the infalling galaxies into the intra-cluster region \citep{gunn72,vollmer01}.
Studying fate of this stripped gas, ionization in contact with million degree hot intra-cluster 
medium, or condensation to form new stars, is of great interest. The Virgo cluster of galaxies 
(hereafter, the Virgo cluster) is the nearest massive cluster of galaxies (16.5~Mpc; 
\citealt{mei07}), and is an ideal laboratory to investigate details of such processes.
We focus on IC~3418, a galaxy likely falling into 
the Virgo cluster for the first time at a very high speed (nearly 1000 km s$^{-1}$; 
\citealt{vollmer01}), especially on its 17~kpc long trail which is considered to have formed 
behind the galaxy due to ram pressure stripping \citep{hester10,fumagalli11}.
The trail is manifested as a chain of blobs seen both in GALEX \citep{martin05} UV continuum 
and H$\alpha$ emission line images \citep{hester10,fumagalli11}. Both of these emissions are clear signs of
recent star formation, and therefore, the trail is the best target to explore 
details of young massive stellar population in the intra-cluster region.
Stellar spectroscopy is unique and useful not only because 
it provides us basic properties of the stellar properties (e.g., age, metallicity, luminosity) 
but also because we can explore a potentially new mode of star formation in this special condition.
We expect that such star formation is quite 
different from those in Milky Way,
and such study will eventually contribute to understanding of galaxy evolution under cluster 
environment \citep{gunn72} and intra-cluster stellar population.

In this letter, we report on our spectroscopic studies of SDSS J122952.66+112227.8 (hereafter, 
SDSSJ1229+1122), a compact source of optical emission within the star-forming trail of IC~3418.

\section{Spectroscopic Observation and Photometry on Archival Images}

FOCAS spectrograph at Subaru telescope \citep{kashikawa02} was used, on the night of 
27 April 2011, in multi-slit spectroscopy mode to study the nature of the trail of IC~3418.
One of our slitlets was put on a faint ($g'_{\rm AB}$=23.0 mag) blue compact object catalogued 
earlier as SDSSJ1229+1122 and located near the far end of the trail (Figures 1 and 2).
Low-resolution spectrum was taken with $0.8''$-wide slit and ``300B'' grism for spectral 
resolution $R$($\equiv \lambda / \delta \lambda$) of $\simeq 700$ to cover 4000\AA--7000\AA, and 
the total exposure time was 6000 s.
Galactic extinction for $E$($B-V$)=0.03 magnitude was corrected based on the NASA Extragalactic Database (NED) information.
Absolute flux scale of the spectrum was re-calibrated based on our photometric data (see below), 
and a small slit-loss correction (by 7\%) was applied. 
Additionally, a shorter exposure (3600 s) but with relatively higher (``medium'') spectral resolution 
($R\simeq 3750$) spectrum was obtained with the same slit width and ``VPH680'' grism, resulting in 
full width at half maximum (hereafter, FWHM) velocity resolution of $\simeq 80$ km s$^{-1}$. 
Low-resolution spectrum is shown in Figure 3, and measured spectroscopic properties on both spectra are summarized in Table 1.

We also performed photometric analysis on archival images to supplement our spectroscopy (Figures 1 and 2).
Canada-France-Hawaii Telescope (CFHT) Megacam \citep{boulade03} images were taken from Megapipe 
\citep{gywn08} at $g'$, $i'$, and $z'$ bands, and another image at $u^{\rm *}$ band was locally 
stacked based on individually pre-calibrated archival images available at CFHT Science Data Archive.
All those images were originally taken for New Generation Virgo cluster Survey (NGVS) project 
\citep{ferrarese12}. SDSSJ1229+1122 has a faint nearby source within $1''$
from our spectroscopic target source, which was not resolved in SDSS image.
SDSSJ1229+1122 was unresolved in both the $i'$ and $z'$ band images having the best FWHM seeing size of $0.7''$ 
($\simeq 60$~pc at the distance of the Virgo cluster). 
Multi-color photometry was performed through double
Gaussian fitting on the images to properly remove any contamination from the nearby
source, and the same amount of Galactic extinction is corrected as for the spectra.
In addition, H$\alpha$ image was obtained with EFOSC2 attached to NTT telescope on 2011 April 29.
Narrow ``Ha\#692'' filter covering H$\alpha$ within velocity range of $-780$ to 2060 km s$^{-1}$ and broad-band 
``R\#642'' filter are used for stellar continuum subtracted H$\alpha$ line image. Total exposure 
times were 1800 s and 600 s for H$\alpha$ and $R$ band images, respectively.
Flux calibration was performed with a spectroscopic standard star LTT3218.
Our point source detection limit ($3\sigma$) is $1.8 \times 10^{-16}$ erg s$^{-1}$ cm$^{-2}$.
Note that this is an independent image from that presented in \cite{hester10} and \cite{fumagalli11}.
Also, deeper GALEX NUV image (1750\AA--2800\AA) was created by combining the GALEX 
Ultraviolet Virgo Cluster Survey (GUViCS) \citep{boselli11} and the Deep Imaging Survey (DIS) 
\citep{martin05}.
The total integration time is 20672 s.
All the images are shown in Figures 1 and 2, and optical photometry data (Table 1) are overlaid on spectra (Figures 3 and 4).

\section{Results}

The low-resolution spectrum of SDSSJ1229+1122 can be characterized by strong H$\alpha$ (at equivalent width (EW) 
of $-34$\AA) and weak H$\beta$ emission lines on blue continuum.
The H$\alpha$ line is slightly resolved with our medium-resolution spectrum, having FWHM of $\sim 164$ km$^{-1}$ 
(corrected for the instrumental resolution), but further line profile analysis was not possible.
Our spectroscopic H$\alpha$ flux ($5.0 \times 10^{-17}$ erg s$^{-1}$ cm$^{-2}$) is below detection limit of our H$\alpha$ image (Figure 2).
The H$\alpha$ flux is $\lesssim 10$ times fainter than nearby compact H$\alpha$ emitters within the trail (\citealt{hester10,fumagalli11}; see also Figure 2).
Our medium-resolution spectrum shows that the heliocentric velocity is $-99$ km s$^{-1}$, 
which is very close to that of IC~3418 ($38$ km s$^{-1}$; \citealt{gavazzi04}).
Their similar recession velocities as well as spatial association of SDSSJ1229+1122 with the trail of IC~3418 (Figure 1) 
strongly support that they are physically associated to each other within the Virgo cluster.
At a distance of 16.5~Mpc (or $m-M=31.1$; \citealt{mei07}), the 
absolute magnitude ($M_{\rm V}$) is $\sim -8.3$ and H$\alpha$ luminosity is $1.6 \times 10^{36}$ erg s$^{-1}$.
The H$\alpha$ luminosity is about 5 -- 100 times fainter than known intra-cluster H~{\sc ii} regions around NGC~4388 
(\citealt{gerhard02,cortese04}) and about 10 times fainter than ones around VCC~1249/M~49 merging system \citep{arrigoni12} in the Virgo cluster.
In one of these known H~{\sc ii} regions, where detailed spectroscopic analyses were made, only a few O-type stars were 
estimated to be powerful enough for the observed luminosity \citep{gerhard02}.

We find that the emission-line ratios of SDSSJ1229+1122 are unusual for photo- or shock-ionized nebulae.
Other than H$\alpha$ and H$\beta$, no nebular forbidden lines were detected, including ones that are typically 
bright in those nebulae ([O~{\sc iii}]$\lambda$ 5007\AA, [N~{\sc ii}]$\lambda\lambda$ 6548,6583\AA, 
and [S~{\sc ii}]$\lambda\lambda$ 6717,6731\AA) (Figure 3).
Their upper limits were measured assuming the same line widths and redshift of Balmer emission lines (Table 1).
Stringent upper limits on low ionization [N~{\sc ii}] and [S~{\sc ii}] and undetected higher ionization [O~{\sc iii}] lines 
strongly indicate that the emission originates neither from H~{\sc ii} regions nor shock-ionized nebulae \citep{vo87} but from hot stellar atmosphere.
We also note that other emission lines seen typically in Luminous Blue Variable (LBV) stars besides Balmer lines, 
such as [He~{\sc i}]$\lambda 5878$\AA~and [Fe~{\sc II}]$\lambda 5018$\AA, were not detected.

Such a luminous compact object with blue continuum should either be a young stellar association or a single blue (super)giant star.
We first searched a stellar spectrum library \citep{pickles98} for matching the continuum  
(below H$\alpha$ and excluding H$\beta$ emission lines) and the broad-band photometric data, and found that 
A2 supergiant (I) is the best match to the observation (Figure 4).
Similar blue continuum can be reproduced by A5 dwarf (V), but its fainter expected $u^{\rm *}$ flux than the observation as well as 
intrinsically faint luminosity does not agree with the observation (Figure 4).
Then we searched {\it starburst99} models of instantaneous star formation \citep{starburst99} for the same purpose.
Continuous star formation models are not considered since they should always show H~{\sc ii}-region like spectrum, unlike our observation.
We chose a model of 0.4 solar metallicity ($Z=0.008$) based on metallically measurement with other intra-cluster H~{\sc ii} regions in the Virgo cluster \citep{gerhard02,arrigoni12}.
Shape of initial mass function (IMF) can not be examined, and we only describe here cases with IMF slope ($\alpha$)=2.35 and $M_{\rm upper}=100$ $M_{\rm \odot}$ since basic results hold with other IMF parameter sets.
Although the model grid provided by \cite{starburst99} is not fine enough to find statistically satisfactory models for all photometry data, we found three possible burst ages to match the observation: 6~Myr, 16~Myr, and 20~Myr (Figure 4).
We found three solutions because, in addition to main sequence stars that become monotonically redder with age, enhanced red supergiant contribution quickly makes the color redder just around $\sim 10$~Myr (or between 6--16~Myr in the assumed model parameters) when most O-type stars evolve away from the main sequence.
Note that the very similar trend of color change is seen for all different IMF parameter sets.
In summary, both possibilities of a blue supergiant (hereafter, BSG) star and young (6--20 Myr) stellar association are equally good to reproduce the observed blue continuum spectrum and broad-band photometry.

\section{Discussions}

\subsection{SDSSJ1229+1122: A compact young stellar association or a single blue supergiant star?}

In the model of instantaneous starburst, the emission line component of SDSSJ1229+1122 should originate from nebula ionized by young massive stars within the stellar association.
Since EW(H$\alpha$) monotonically decreases with the burst age, only the $\sim 6$~Myr model among the three possibilities can reproduce rather large observed EW(H$\alpha$) ($-34$\AA) due to remaining main sequence O-type stars \citep{starburst99}.
The continuum luminosity then requires stellar mass of $\sim 10^4$ $M_{\rm \odot}$, and expected number of O-type stars there is several -- several 10s depending on choices of the IMF parameters, producing luminous H~{\sc ii} region spectrum that is not observed.
Next possibility is that a stellar association of $\sim 16$--20~Myr dominates the observed continuum and extra H$\alpha$ emitting source(s) in addition to regular population produce strong H$\alpha$.
Possible extra emission-line components are classical Be stars (\citealt{lamers98,paul12}), supergiant Be stars, and supergiant A stars (\citealt{tully84,mccarthy97}).
All of them show prominent H$\alpha$ emission without nebular lines, and their EW(H$\alpha$) alone is already comparable to the observed one (e.g., \citealt{tully84,paul12}), indicating that those stars dominate the observed spectrum since underlying spectrum from regular stellar association, if any, will dilute the observed width.
The observed absolute magnitude ($M_{\rm V}\sim -8.3$) requires either a single BSG or 50--200 classical Be stars (class III--V) \citep{lamers98,paul12} since they are fainter by 4--6 magnitude than BSGs.
If SDSSJ1229+1122 is made entirely of classical Be stars, such larger number of short-lived objects within a compact ($\lesssim 60$~pc across) 
stellar association requires a very unlikely star formation episode.
Instead, if it is a BSG, both the absolute magnitude and EW(H$\alpha$) are within the range of A0~I stars in SMC \citep{tully84} although the observed EW(H$\beta$) is larger by a factor of $\sim 2$.
Since the EW in BSGs is a function of metallicity, luminosity, and stellar type \citep{tully84}, we believe that our observation is 
within the probable range of BSGs given uncertainties of detailed stellar parameters for SDSSJ1229+1122.
Observed intrinsically broad (FWHM of $\sim 164$ km$^{-1}$) H$\alpha$ emission hints presence of strong stellar wind and/or fast rotating equatorial disk, although their characteristic profile (e.g., P-Cygni type for BSGs; \citealt{mccarthy97,prybilla06}) could not be discerned.
Therefore a model of single BSG as the optically dominant sources within SDSSJ1229+1122 is strongly preferred over a model of young stellar association. Note, however, that we can not distinguish between case of a single star or collection of a few such stars based on our observation, although only a single such star is bright enough.
We expect that SDSSJ1229+1122 was an H~{\sc ii} region about a 
few -- several Myr ago and similar in nature to other known H~{\sc ii} regions in the trail of IC~3418 \citep{hester10,fumagalli11} and ones within the Virgo cluster \citep{gerhard02,cortese04,arrigoni12}.

\subsection{Star formation characteristics within the cluster environment}

Stars usually form in dusty giant molecular gas clouds, but such clouds are too tightly bound 
with the galaxy to be stripped off by ram pressure stripping and thrown into the gaseous trail 
of cluster-infalling galaxies \citep{hester10}.
However, a few rare trails including one of IC~3418 and others that extend up to 90~kpc 
from the parent galaxy are seen decorated with UV-bright, H$\alpha$ emitting, and/or 
optically-blue young star-forming clumps (\citealt{cortese07,yoshida08,hester10,fumagalli11,arrigoni12}).
\cite{fumagalli11} analyzed stellar properties of IC~3418 based on optical and GALEX-UV spectral energy distribution 
(SED), and argue that the star formation there has been 
truncated about 200~Myr ago due to ram pressure stripping.
As massive stars are short-lived, the O-type star, which has evolved into a BSG by now, must have formed a few 10~ Myr ago within SDSSJ1229+1122, i.e., it formed after the truncation of star formation within IC~3418.
This supports an idea that the star have formed in situ within the trail.

SDSSJ1229+1122 is located within a diffuse GALEX NUV blob, labeled ``D3'' by \cite{hester10} or a ``lower surface-brightness and visually more diffuse filament'', labeled ``F1'' by \cite{fumagalli11}.
The blob extends $\sim 10''$ or $\sim 800$~pc.
\cite{fumagalli11} reported 2.3 magnitude brighter $g'$ band magnitude for F1 than our measurement on SDSSJ1229+1122, and analyzed that the overall SED including GALEX-UV bands can be reproduced by young stellar population with age of 640$^{+248}_{-144}$~Myr and mass of $3.48 \times 10^5$ $M_{\rm \odot}$.
However, the CFHT image shows only a few more fainter compact sources besides SDSSJ1229+1122 there (Figure 2), indicating that there is an unresolved diffuse young stellar population within the blob.
We speculate that there are much more unexplored population of blue 
stellar population within the intra-cluster region, not only along the trail of IC~3418 but 
also in other similar trails in the Virgo cluster.

It is likely that star formation in such a special environment is quite different from typical 
cases in, e.g., our own Milky Way galaxy. Difference in temperature and relative velocity 
between the star-forming molecular clouds and the ambient medium may be million degree and $\sim 1000$ km s$^{-1}$, respectively.
Being far from the parent galaxy, especially at the far end of the trail, role of turbulence may be dominant over gravity.
Turbulence in the trail may create eddies to form dense cloud clumps 
which would cool very fast and subsequently collapse under their own gravity to form stars.
Although the model of turbulence-driven star formation seems promising 
\citep{hester10,fumagalli11}, constraining detailed physical mechanism of the star formation 
and characterizing the turbulence requires further investigations of stellar population and cool molecular gas there.

\section{Implication for Future Extragalactic Stellar Astrophysical Studies}

Various kinds of intra-cluster stellar population have been known in the Virgo cluster:
In addition to diffuse intra-cluster optical continuum light \citep{mihos05}, individual 
sources like intra-cluster H~{\sc ii} regions (\citealt{gerhard02,cortese04,arrigoni12}), 
old ($\ge 10$ Gyr) intra-cluster stars \citep{williams07}, and 
intra-cluster planetary nebulae (\citealt{feldmeier04,aguerri05}) are known.
Our study, independent of the interpretation that SDSSJ1229+1122 contains a single or a few more stars, has added BSGs to be residing in the intra-cluster region.
Previously, BSG candidates have been photometrically identified in M~100 \citep{hill98} in the Virgo cluster, and spectroscopically identified in NGC 3621 at a distance of 6.7 Mpc (Bresolin et al. 2001). 
Being in the Virgo cluster, and if confirmed by future observations to be a single star, SDSSJ1229+1122 will be the most distant star discovered from spectroscopic observations.
We demonstrated for the first time that stellar spectroscopy up to distance of the Virgo cluster is indeed feasible as anticipated nearly two decade ago (\citealt{kudritzki95,mccarthy97,bresolin01,kudritzki10}).
However, to reveal its true nature by detailed quantitative spectroscopic analyses, spectra with higher signal-to-noise ratio and higher resolution below H$\beta$ ($\le 4800$\AA) as well as H$\alpha$ are required.
To exploit its full potential for accurate distance measurements and star formation studies in the Virgo cluster and beyond, we do need the next-generation giant telescopes such as Thirty Meter Telescope (TMT) and/or European Extremely Large Telescope (E-ELT) \citep{kudritzki10}.

\acknowledgments

Based in part on data collected at Subaru Telescope, which is operated by the National Astronomical Observatory of Japan.
Based on observations obtained with MegaPrime/MegaCam, a joint project of CFHT and CEA/DAPNIA, at the Canada-France-Hawaii Telescope (CFHT) which is operated by the National Research Council (NRC) of Canada, the Institut National des Science de l'Univers of the Centre National de la Recherche Scientifique (CNRS) of France, and the University of Hawaii.
This research used the facilities of the Canadian Astronomy Data Centre operated by the National Research Council of Canada with the support of the Canadian Space Agency.
Funding for the SDSS and SDSS-II has been provided by the Alfred P. Sloan Foundation, the Participating Institutions, the National Science Foundation, the U.S. Department of Energy, the National Aeronautics and Space Administration, the Japanese Monbukagakusho, the Max Planck Society, and the Higher Education Funding Council for England.
Based on data obtained from the ESO Science Archive Facility under request number ``Youichi Ohyama \#35,191''.
Some of the data presented in this paper were obtained from the Mikulski Archive for Space Telescopes (MAST).
This research has made use of the NASA/IPAC Extragalactic Database (NED) which is operated by the Jet Propulsion Laboratory, California Institute of Technology, under contract with the National Aeronautics and Space Administration. 
This work is supported by grant NSC 100-2112-M-001-001-MY3 (Y.O.). 
A.H. is thankful to Inter-University Centre for Astronomy and Astrophysics (IUCAA), India for a visit with financial support and 
to National Centre for Radio Astrophysics of Tata Institute of Fundamental Research (NCRA-TIFR), India for a 
Visiting Astronomer position during which parts of the work were done. 

\newpage

\begin{figure*}
\epsscale{1}
\begin{center}
\includegraphics[clip=false,scale=0.6,angle=-90]{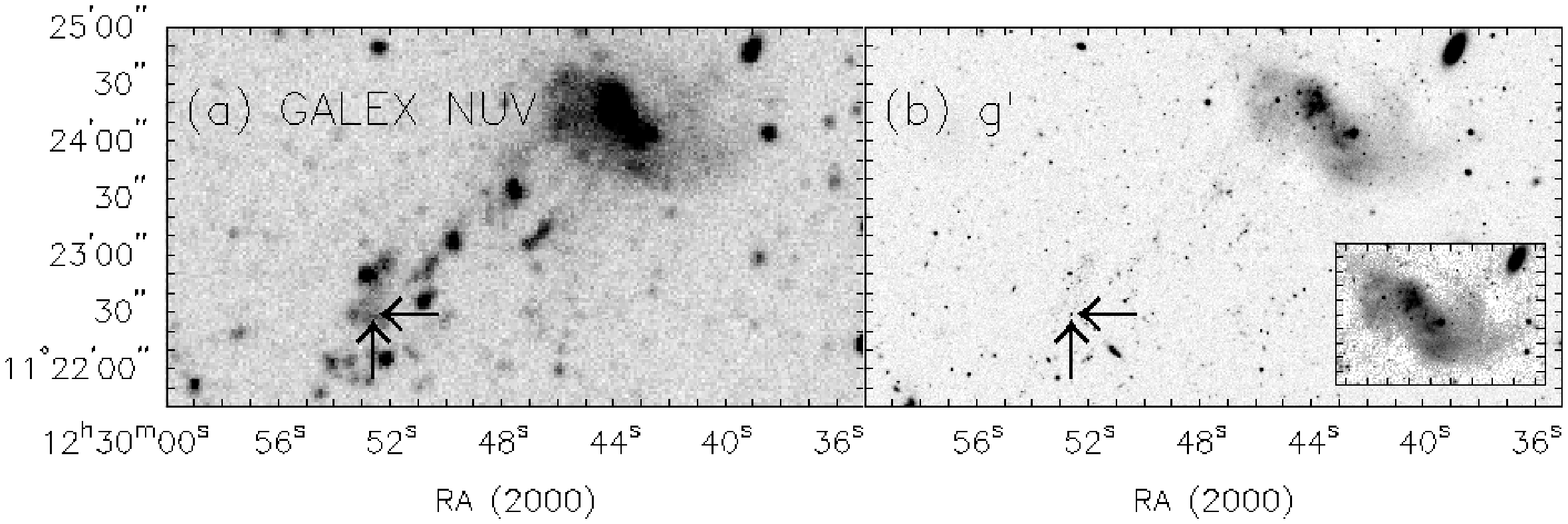}
\end{center}
\caption{
GALEX NUV (a) and CFHT $g'$ band (b) images of IC~3418 and its trail including SDSSJ1229+1122 (marked by two arrows) in linear intensity scale.
An inset $g'$ band image within (b) shows fainter tidal features of IC~3418 in square-root intensity scale.
}
\end{figure*}

\newpage

\begin{figure*}
\epsscale{1}
\begin{center}
\includegraphics[clip=false,scale=0.6,angle=-90]{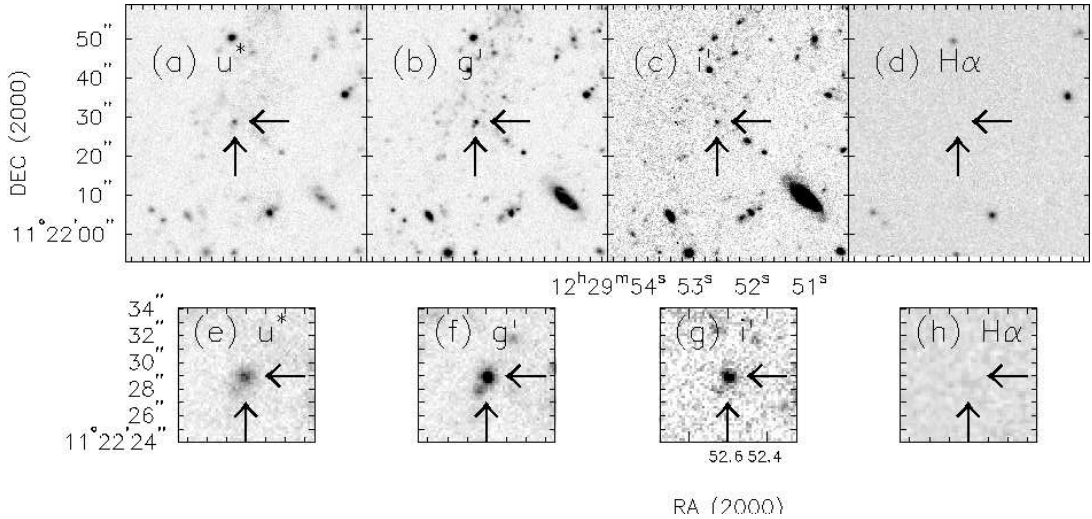}
\end{center}
\caption{
Zoomed-in optical views around SDSSJ1229+1122.
Upper panels show $\sim 1'$ view around SDSSJ1229+1122 in $u^{\rm *}$, $g'$, $i'$, and continuum-subtracted H$\alpha$ line ((a) -- (d)).
Lower panels show the same images but just for central $\sim 10''$ region ((e) -- (h)).
All images are shown in linear intensity scale.
}
\end{figure*}

\newpage

\begin{figure*}
\begin{center}
\includegraphics[clip=false,scale=0.9,angle=90]{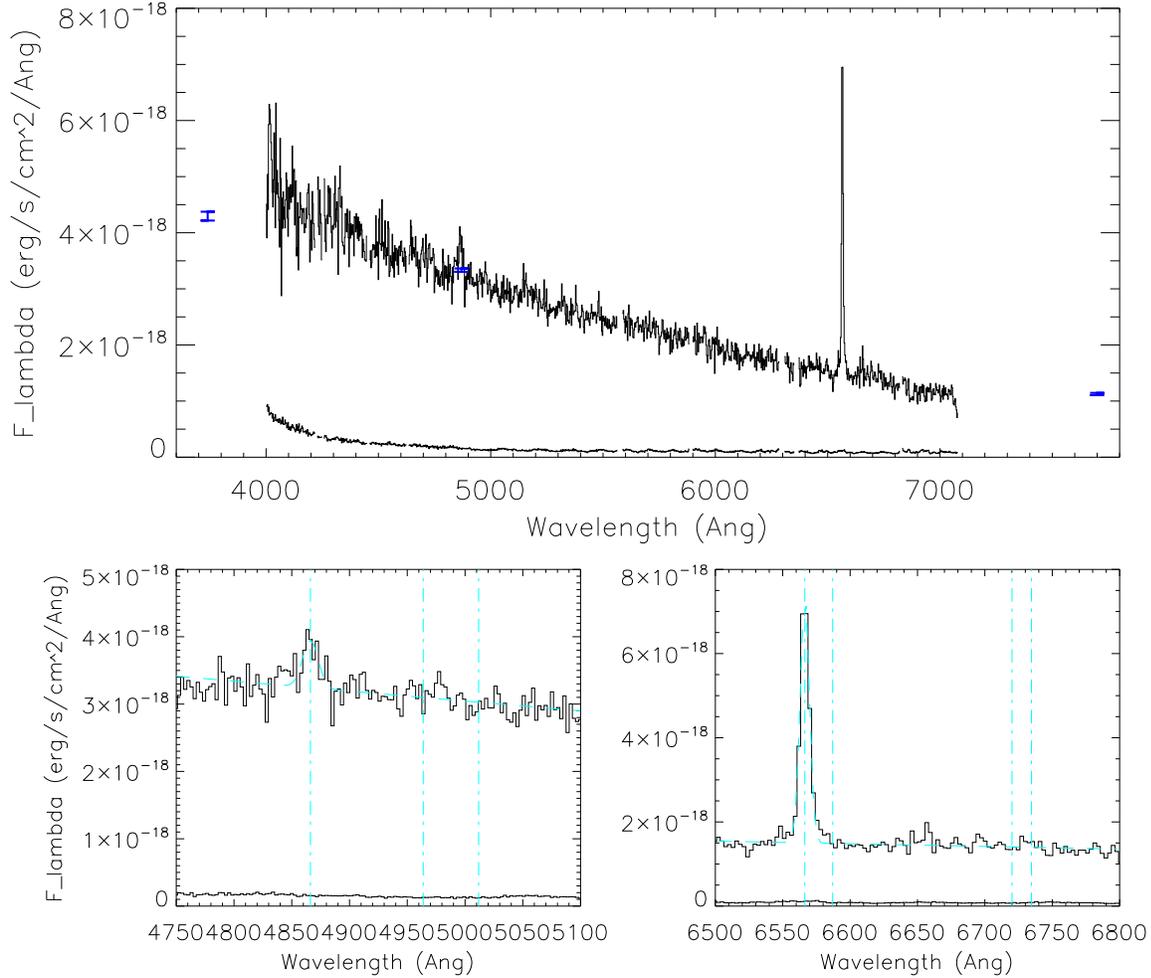}
\end{center}
\caption{
{\it Top panel:} Low-resolution optical spectrum and $u^{\rm *}$, $g'$, and $i'$ band photometry ({\it blue} with error bars) of SDSSJ1229+1122.
One-sigma noise spectrum
is shown near the bottom of all plots.
{\it bottom left:} The same spectrum of {\it top} panel but around H$\beta$ and [O~{\sc iii}].
Multi-Gaussian fitting (plus linear function for the continuum) results and the line wavelengths are overlaid with {\it cyan} broken lines.
Neither [O~{\sc iii}]$\lambda 4959$ nor $\lambda 5007$ are detected but their wavelengths expected for the redshift of H$\beta$ are marked.
{\it bottom right:} The same spectrum of {\it top} but around H$\alpha$, [N~{\sc ii}], and [S~{\sc ii}].
Multi-Gaussian fitting results are shown in the same way as for the {\it bottom left}.
Neither [N~{\sc ii}]$\lambda 6583$, [S~{\sc ii}]$\lambda 6717$, nor $\lambda 6731$ are detected but their wavelengths expected for the redshift of H$\alpha$ are marked.
}
\end{figure*}

\newpage

\begin{figure*}
\begin{center}
\includegraphics[clip=false,scale=0.9]{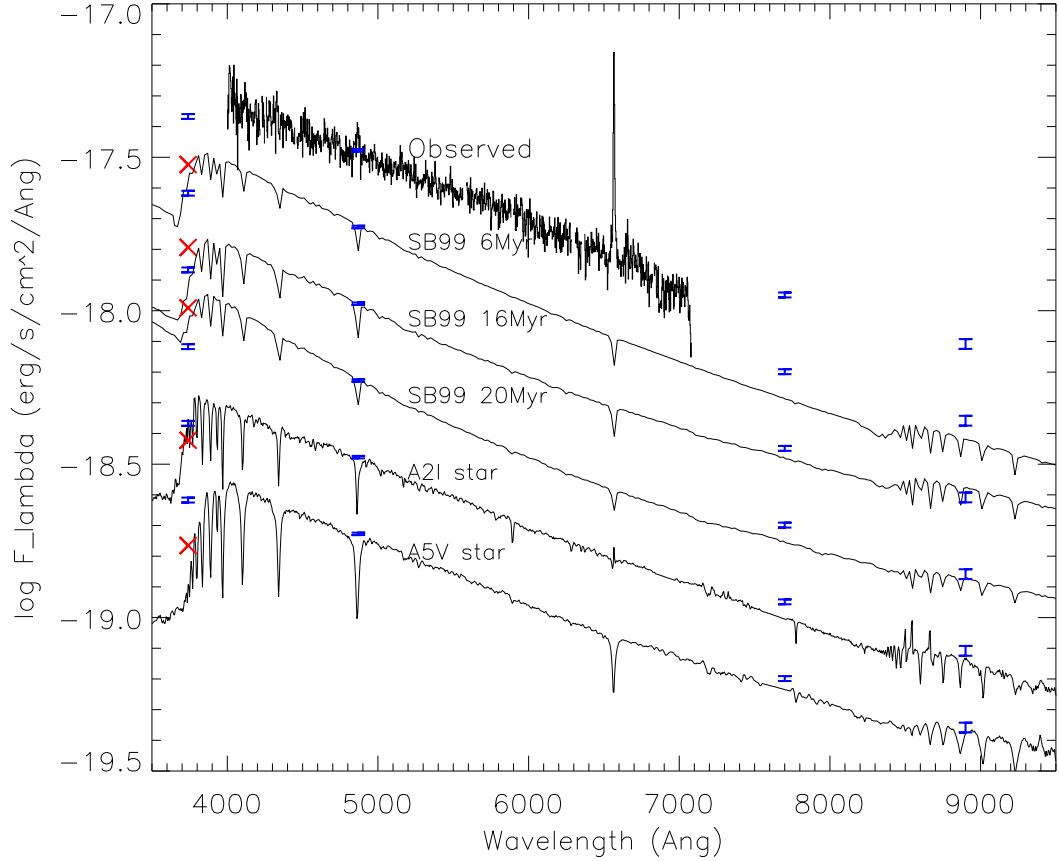}
\end{center}
\caption{
Comparison of observed low-resolution spectrum and photometry data for $u^{\rm *}$, $g'$, $i'$, and $z'$ bands 
({\it blue} with error bars) of SDSSJ1229+1122 ({\it top}) with model spectra.
Model spectra are scaled at $g'$ band to the observation, and then offset by $-0.25$ in log flux scale along with the 
photometry data for clarity of the figure.
As labeled, second, third, and fourth spectra 
are {\it starburst99} instantaneous models of 6~Myr, 16~Myr, and 20~Myr, and following two spectra are of A2~I and A5~V stars, respectively.
Expected $u^{\rm *}$ fluxes for each model are marked with {\it red} crosses.
}
\end{figure*}

\newpage

\begin{table}
\caption{Photometric and spectroscopic characteristics of SDSSJ1229+1122}
\begin{tabular}{lcc}
\tableline
\tableline
Megacam astrometry & RA (2000) & DEC (2000) \\
 & 12$^{\rm h}$29$^{\rm m}$52$^{\rm s}$.69 & +11$^{\rm \circ}$22$'$28.0$''$ \\
\tableline
Megacam photometry\tablenotemark{1} & Observed & Corrected\tablenotemark{2} \\
$u^{\rm *}$ & $23.29\pm 0.02$ & $23.15\pm 0.02$ \\
$g'$ & $22.96\pm 0.01$ & $22.85\pm 0.01$ \\
$i'$ & $23.09\pm 0.02$ & $23.03\pm 0.02$ \\
$z'$ & $23.16\pm 0.04$ & $23.12\pm 0.04$ \\
Bessel photometry\tablenotemark{2,3} & Apparent & Absolute\tablenotemark{4} \\
$V$ & $22.85\pm 0.01$ & $M_{\rm V}=-8.25$ \\
\tableline
FOCAS low-resolution spectroscopy & & \\
H$\alpha$ flux\tablenotemark{2} & \multicolumn{2}{c}{($5.04\pm 0.08$) $\times 10^{-17}$ erg s$^{-1}$ cm$^{-2}$} \\
H$\alpha$ equivalent width & \multicolumn{2}{c}{$-33.7\pm 0.7$ \AA} \\
H$\beta$ flux\tablenotemark{2} & \multicolumn{2}{c}{($2.08\pm 0.22$) $\times 10^{-17}$ erg s$^{-1}$ cm$^{-2}$} \\
H$\beta$ equivalent width & \multicolumn{2}{c}{$-6.7\pm 0.7$ \AA} \\
\tableline
Emission line ratios & \multicolumn{1}{c}{$3\sigma$ upper limit} & \\
$\rm [O III]\lambda$5007/H$\beta$ & \multicolumn{2}{c}{$< 0.28$} \\
$\rm [N II]\lambda$6583/H$\alpha$ & \multicolumn{2}{c}{$< 0.028$} \\
$\rm [S II]\lambda$6717+6731/H$\alpha$ & \multicolumn{2}{c}{$< 0.031$} \\
\tableline
FOCAS medium-resolution spectroscopy & & \\
H$\alpha$ heliocentric velocity & \multicolumn{2}{c}{$-99\pm 6$ km s$^{-1}$} \\
H$\alpha$ intrinsic line width & \multicolumn{2}{c}{$164\pm 22$ km s$^{-1}$ FWHM} \\
\tableline
\end{tabular}

\tablenotetext{1}{In Megacam SDSS ($AB$) system.}
\tablenotetext{2}{Galactic extinction corrected for $E$($B-V$)=0.03.}
\tablenotetext{3}{Bessel system in Vega unit, converted based on our spectrum.}
\tablenotetext{4}{Distance modulus of $m-M=31.1$ is used.}

\end{table}

\end{document}